\documentclass[preprint,article,accept,moreauthors,pdftex]{Definitions/mdpi} 

\firstpage{1} 
\pubvolume{xx}
\issuenum{1}
\articlenumber{5}
\pubyear{2021}
\copyrightyear{2021}
\history{Received: date; Accepted: date; Published: date}
\pdfoutput=1

\usepackage{xcolor}
\usepackage{physics}
\usepackage{subfiles}
\usepackage{subcaption}

\Title{An investigation into the approximations used in wave packet molecular dynamics for the study of warm dense matter}

\Author{William A. Angermeier$^{1}$, Thomas G. White$^{1,*}$ }

\AuthorNames{William A. Angermeier and Thomas G. White}

\address{%
$^{1}$ \quad University of Nevada, Reno, Department of Physics; tgwhite@unr.edu\\
}

\corres{Correspondence:tgwhite@unr.edu (T.G.W.)}

\abstract{
Wave packet molecular dynamics (WPMD) has recently received a lot of attention as a computationally fast tool to study dynamical processes in warm dense matter beyond the Born-Oppenheimer approximation. These techniques, typically, employ many approximations to achieve computational efficiency while implementing semi-empirical scaling parameters to retain accuracy. We investigate three of the main approximations ubiquitous to WPMD: a restricted basis set, approximations to exchange, and the lack of correlation. We examine each of these approximations in atomic and molecular hydrogen in addition to a dense hydrogen plasma. We find that the biggest improvement to WPMD comes from combining a two Gaussian basis with a semi-empirical correction based on the valence-bond wave function. A single parameter scales this correction to match experimental pressures of dense hydrogen. Ultimately, we find that semi-empirical scaling parameters are necessary to correct for the main approximations in WPMD. However, reducing the scaling parameters for more ab-initio terms gives more accurate results and displays the underlying physics more readily.
}


\begin{document}

%%%%%%%%%%%%%%%%%%%%%%%%%%%%%%%%%%%%%%%%%%%%%%%%%%%%%%%%%%%%%%%%%%%%%%%%%%%%%%%%%%%%%%%%%%%%%%%%%%%%%%%%%%%%%%%%%%%%%%%%%%%%%%%%%%%%%%%%%%%%%%%%%%%%%%%%%%%%%
%%%%%%%%%%%%%%%%%%%%%%%%%%%%%%%%%%%%%%%%%%%%%%%%%%%%%%%%%%%%%%%%%%%%%%%%%%%%%%%%%%%%%%%%%%%%%%%%%%%%%%%%%%%%%%%%%%%%%%%%%%%%%%%%%%%%%%%%%%%%%%%%%%%%%%%%%%%%%

\section{Introduction}

%%%%%%%%%%%%%%%%%%%%%%%%%%%%%%%%%%%%%%%%%%%%%%%%%%%%%%%%%%%%%%%%%%%%%%%%%%%%%%%%%%%%%%%%%%%%%%%%%%%%%%%%%%%%%%%%%%%%%%%%%%%%%%%%%%%%%%%%%%%%%
%%%%%%%%%%%%%%%%%%%%%%%%%%%%%%%%%%%%%%%%%%%%%%%%%%%%%%%%%%%%%%%%%%%%%%%%%%%%%%%%%%%%%%%%%%%%%%%%%%%%%%%%%%%%%%%%%%%%%%%%%%%%%%%%%%%%%%%%%%%%%
%%%%%%%%%%%%%%%%%%%%%%%%%%%%%%%%%%%%%%%%%%%%%%%%%%%%%%%%%%%%%%%%%%%%%%%%%%%%%%%%%%%%%%%%%%%%%%%%%%%%%%%%%%%%%%%%%%%%%%%%%%%%%%%%%%%%%%%%%%%%%
Warm Dense Matter (WDM) is a critically important physical regime that bridges the gap between condensed matter and classical plasma physics. The WDM state is found in several astrophysical environments (e.g., planetary interiors and white dwarfs) \cite{guillot1999, paquette}.  It also has practical applications for understanding controlled thermonuclear fusion, and material processing \cite{atzeni}. Typically described as a system of strongly coupled ions immersed in a degenerate electron sea, WDM may exist in either a compressed liquid or a highly excited solid state. In both states, the ions have a Coulomb energy comparable to the thermal energy, while the electrons, at temperatures below the Fermi temperature,  exhibit strong quantum behavior \cite{ichimaru1982}. Techniques that simulate WDM states must model the slow and long-time behavior of the strongly coupled ions while simultaneously capturing the electrons' quantum mechanical nature. These inherent complexities lead to the failure of perturbative techniques resulting in differences in predictions of important quantities. While different models generally agree on the thermodynamic and acoustic properties \cite{davis2020,yao2021}, important quantities such as transport coefficients can differ by up to an order-of-magnitude \cite{lorenzen,grabowski2020}.

%%%%%%%%%%%%%%%%%%%%%%%%%%%%%%%%%%%%%%%%%%%%%%%%%%%%%%%%%%%%%%%%%%%%%%%%%%%%%%%%%%%%%%%%%%%%%%%%%%%%%%%%%%%%%%%%%%%%%%%%%%%%%%%%%%%%%%%%%%%%%
%%%%%%%%%%%%%%%%%%%%%%%%%%%%%%%%%%%%%%%%%%%%%%%%%%%%%%%%%%%%%%%%%%%%%%%%%%%%%%%%%%%%%%%%%%%%%%%%%%%%%%%%%%%%%%%%%%%%%%%%%%%%%%%%%%%%%%%%%%%%%
%%%%%%%%%%%%%%%%%%%%%%%%%%%%%%%%%%%%%%%%%%%%%%%%%%%%%%%%%%%%%%%%%%%%%%%%%%%%%%%%%%%%%%%%%%%%%%%%%%%%%%%%%%%%%%%%%%%%%%%%%%%%%%%%%%%%%%%%%%%%%
Atomistic models in which the ions, treated through classical molecular dynamics, are coupled with a quantum mechanical treatment of the electrons have had the most success. The ion trajectories from such simulations can provide transport properties, such as viscosity and thermal diffusivity \cite{salin2003}, acoustic properties, such as the sound speed \cite{white2013,mabey2017}, and thermodynamic variables, including the equation of state \cite{graziani2012,white2013}. The most prevalent of these techniques is density functional theory molecular dynamics (DFT-MD), in which the electrons are treated within the framework of either orbital-free \cite{white2013} or Kohn-Sham density functional theory \cite{ruter2014}. DFT-MD employs the Born-Oppenheimer (BO) approximation, in which the electrons are considered to respond instantaneously to the ion dynamics, usually justified by the disparate time-scales of the electron and ion motion. Although applicable for equilibrium properties, such as the equation-of-state \cite{kang2020}, the BO treatment of the ions may not be suitable for calculating dynamic properties such as sound-speed and transport coefficients \cite{mabey2017,yao2021}. Furthermore, by its very nature, the BO approximation prohibits direct energy transfer between electrons and ions and is therefore problematic for modeling non-equilibrium matter \cite{clerouin2015}. Ultimately, in the WDM regime, it is still unknown how the BO approximation impacts atomistic calculations.

Recently, several theoretical techniques have been developed that go beyond the BO approximation. The simplest technique couples the ions to a  Langevin thermostat which models the electron-ion collisions through an additional stochastic Gaussian force added to the equations-of-motion \cite{mabey2017, dai2009, tamm2018, duffy2007,rutherford2007}. While efficient, this phenomenological approach uses a single collision frequency that must be determined \emph{a priori}. Other techniques that go beyond the BO approximation include using a thermally averaged, linearized Bohm potential, which has successfully described acoustic oscillations in warm dense aluminum  \cite{larder2019, moldabekov2021}, and the most widespread, Wave Packet Molecular Dynamics (WPMD) \cite{klakow1994,knaup2001,jakob2007, su2007,lavrinenko2018,davis2020,yao2021,jaramillo-botero2011,kim2011,ma2019}. Time-dependent DFT, often used successfully in quantum chemistry \cite{li2005}, remains too computationally intensive to study the large systems of interest. In this work, we will focus on the applicability and approximations used within WPMD.

%%%%%%%%%%%%%%%%%%%%%%%%%%%%%%%%%%%%%%%%%%%%%%%%%%%%%%%%%%%%%%%%%%%%%%%%%%%%%%%%%%%%%%%%%%%%%%%%%%%%%%%%%%%%%%%%%%%%%%%%%%%%%%%%%%%%%%%%%%%%%
%%%%%%%%%%%%%%%%%%%%%%%%%%%%%%%%%%%%%%%%%%%%%%%%%%%%%%%%%%%%%%%%%%%%%%%%%%%%%%%%%%%%%%%%%%%%%%%%%%%%%%%%%%%%%%%%%%%%%%%%%%%%%%%%%%%%%%%%%%%%%
%%%%%%%%%%%%%%%%%%%%%%%%%%%%%%%%%%%%%%%%%%%%%%%%%%%%%%%%%%%%%%%%%%%%%%%%%%%%%%%%%%%%%%%%%%%%%%%%%%%%%%%%%%%%%%%%%%%%%%%%%%%%%%%%%%%%%%%%%%%%%
WPMD is a time-dependent quantum mechanical technique that simultaneously simulates, (1) the propagation of the ions as classical point particles, and (2) the electrons as quantum mechanical entities. In WPMD, each electron is represented as a quantum wave-packet, a spatially-localized complex function often implemented on a Gaussian basis \cite{feldmeier2000}. These wave-packets uniquely define the state of a single electron, with the total many-body wave function constructed from either a Hartree product or Slater determinant \cite{su2009, grabowskiReviewWP}. A choice that is driven by the importance of balancing exchange effects with computational cost. Equations-of-motion for the dynamical parameters are easily derived from variation of the time-dependent Schrodinger equation, where, for a single Gaussian basis, they take on a simple Hamilton form \cite{feldmeier2000, su2007}. The direct inclusion of electrons, and thus the effects of electron-ion interactions, means that WPMD intrinsically goes beyond the BO approximation. It is capable of computing electron-ion energy exchange in non-equilibrium systems, the effects of electron-ion collisions, and more generally calculating observables in quantum many-body systems \cite{ma2019, grabowskiReviewWP}. 

Many flavors of WPMD exist that utilize varying degrees of approximation. The three most common approximations are a restricted basis, consisting of a single Gaussian per electron, a pairwise exchange interaction, often identified as a Pauli potential, and the exclusion of correlation. Furthermore, the Pauli potential itself is often assumed to depend only on the kinetic energy component of exchange in addition to the dependence on electron momentum being ignored \cite{boal1988,klakow1994,su2007}. With these simplifications, WPMD can obtain the same computational efficiency as in many classical methods \cite{morozov2009,grabowski2013}. These efficiencies have led to the widespread use of a semi-empirical WPMD method known as the electron force field (eFF). In eFF, several experimentally derived scaling parameters are used, with remarkable success, to correct for deficiencies in basis, approximations to exchange, and lack of correlation. To date, eFF has been used to investigate material properties in extreme environments \cite{su2009,xiao2015,jaramillo-botero2011,kim2011,lan2020}, temperature relaxation rates in warm dense hydrogen \cite{davis2020}, sound-speed in warm dense aluminum \cite{davis2020}, and diffusion in warm dense hydrogen \cite{yao2021}. 

\begin{figure}[H]
    \centering
    \includegraphics[width=9cm, keepaspectratio]{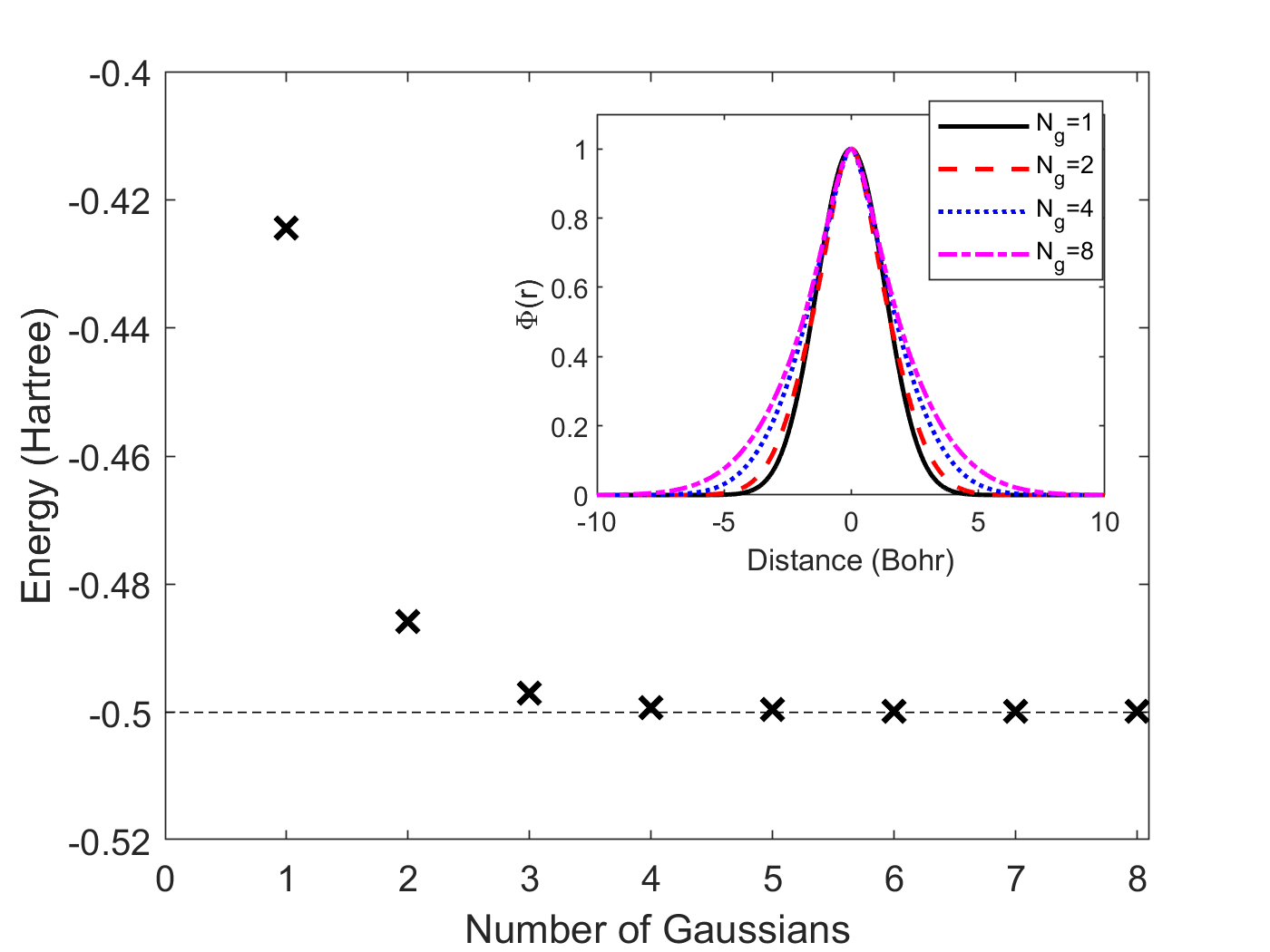}
    \caption{\textbf{The ground state energy of the hydrogen atom} The crosses are the calculated from minimization of the energy equations as the number of Gaussian basis functions are increased. The dotted line shows the exact ground state energy of hydrogen \cite{cohen-tannoudji}. The insert shows the changes to the shape of the wave functions with an increasing number of basis functions. Beyond four Gaussians we find negligible improvement in the either the ground state energy or shape of the wavefunction.}
    \label{fig:Hatom}
\end{figure}

Despite the success of eFF, exactly how the scaling parameters address these approximations is not well understood. Predictive capability is limited, and its use in new regimes should always be corroborated with other models or experimental data \cite{grabowskiReviewWP}. For example, eFF was recently shown to underestimate ion-ion correlation in dense aluminum plasmas at temperatures of a few electronvolts \cite{davis2020,fletcher2015}. While there has been some effort to understand the effect of a pair-wise exchange in dense hydrogen \cite{klakow1994,knaup2001,knaup2002,knaup2003,jakob2007,su2007}, the basis set limitation has not previously been investigated. However, work involving a multiple-Gaussian basis has been used to investigate wave-packet spreading in electron-nuclear scattering \cite{grabowskiReviewWP} and ionization of a single hydrogen atom \cite{morozov2012,valuev2015}; where improvements up to a five Gaussian basis were found. Figure \ref{fig:Hatom} shows improvement to the ground state energy of the hydrogen atom with an increasing number of Gaussians in the electron basis; minimal improvement is observed beyond four Gaussians.  

We will focus our efforts on dense hydrogen, the prototypical test-bed for atomistic models, and investigate the accuracy of WPMD as the number of Gaussians in the basis is increased. For computational efficiency, we retain a pairwise Pauli potential; however, based on the work of H. Xiao \cite{xiaothesis2015}, we extend the Pauli potential to include additional potential energy terms. Finally, a simple correction based on the Valence Bond (VB) wave function is introduced with a single scaling parameter to address the model's lack of correlation and pairwise exchange. With this correction and a two-gaussian basis, we are able to exactly match the low-temperature pressure curve of dense hydrogen. Ultimately, we hope to extend the parameter space where WPMD, and specifically eFF, is applicable.

%%%%%%%%%%%%%%%%%%%%%%%%%%%%%%%%%%%%%%%%%%%%%%%%%%%%%%%%%%%%%%%%%%%%%%%%%%%%%%%%%%%%%%%%%%%%%%%%%%%%%%%%%%%%%%%%%%%%%%%%%%%%%%%%%%%%%%%%%%%%%
%%%%%%%%%%%%%%%%%%%%%%%%%%%%%%%%%%%%%%%%%%%%%%%%%%%%%%%%%%%%%%%%%%%%%%%%%%%%%%%%%%%%%%%%%%%%%%%%%%%%%%%%%%%%%%%%%%%%%%%%%%%%%%%%%%%%%%%%%%%%%
%%%%%%%%%%%%%%%%%%%%%%%%%%%%%%%%%%%%%%%%%%%%%%%%%%%%%%%%%%%%%%%%%%%%%%%%%%%%%%%%%%%%%%%%%%%%%%%%%%%%%%%%%%%%%%%%%%%%%%%%%%%%%%%%%%%%%%%%%%%%%

This paper is organized as follows. Section 2 details the theory and development of the WPMD equations for systems with a multiple Gaussian basis. Particular focus will be given to the development of the updated Pauli potential, including a comparison of the updated Pauli potential with past work. Section 3 details the improvements afforded to hydrogen-based systems. The results include geometry optimization, potential energy surfaces and dynamics of the hydrogen molecule, and hydrogen under high pressure. Where available, we benchmark our calculations to experimental results. Finally, section 4 details some of the effects of approximations in current WPMD techniques and suggests a path forward to provide large-scale simulations of dense plasmas in previously unexplored regions of phase space. 

%%%%%%%%%%%%%%%%%%%%%%%%%%%%%%%%%%%%%%%%%%%%%%%%%%%%%%%%%%%%%%%%%%%%%%%%%%%%%%%%%%%%%%%%%%%%%%%%%%%%%%%%%%%%%%%%%%%%%%%%%%%%%%%%%%%%%%%%%%%%%%%%%%%%%%%%%%%%%
%%%%%%%%%%%%%%%%%%%%%%%%%%%%%%%%%%%%%%%%%%%%%%%%%%%%%%%%%%%%%%%%%%%%%%%%%%%%%%%%%%%%%%%%%%%%%%%%%%%%%%%%%%%%%%%%%%%%%%%%%%%%%%%%%%%%%%%%%%%%%%%%%%%%%%%%%%%%%

\section{Materials and Methods}

\subsection{Introduction to Wave-packet Molecular Dynamics}

In WPMD the many-electron trial wave function $\Psi$ is parameterized by a set of variables $(\mathbf{q}(t))$ \cite{klakow1994,knaup2001,su2007,jakob2007,morozov2012}. Development of a fully anti-symmetric wave function requires a calculation of the Slater determinant of all single electron wave functions, a computationally expensive operation \cite{jakob2007, morozov2012, valuev2015}. To decrease computational cost, many implementations of WPMD, including eFF, use the simpler Hartree product, constructed through the linear combination of the single-particle wave-functions, $\psi_{k}$. The Hartree product wave function, $\Psi^{H}$, is defined as:

\begin{equation}
     \Psi^{H}(\vec{X}, t) = \prod_{k=1}^{N_{e}} \psi_{k}(\vec{x}_{k}, t),
     \label{Eq:Hartree}
\end{equation} 
where $\vec{X} = \{ \vec{x}_{1}, ..., \vec{x}_{k}, ..., \vec{x}_{N_{e}} \}$, and $\vec{x}_k$ are used to indicate the space occupied by the k-th electron. Most WPMD techniques utilize a single isotropic Gaussian as a restricted wave function for each electron \cite{boal1988,klakow1994,knaup2001,jakob2007,su2009}. This wave function is typically paramaterized by a set of ten real physical variables $\mathbf{q}=\{\vec{r}, \vec{p}, \sigma, p_\sigma\}$ i.e.,

\begin{align}
\psi_{k}(\vec{x}_{k})=\left(\frac{3}{2 \pi \sigma^{2}}\right)^{3 / 4} \exp \left[-\left(\frac{3}{4 \sigma^{2}}-\frac{i p_{\sigma}}{2 \hbar \sigma}\right)\left|\vec{r}-\vec{x}_{k}\right|^{2}+\frac{i \vec{p} \cdot\left(\vec{r}-\vec{x}_{k}\right)}{\hbar}\right].
\label{Eq:GaussianPhys}
\end{align} 
The elegance of this definition is that $\vec{r}=\langle \hat{r} \rangle$ is the expectation of position, $\vec{p}=\langle \hat{p} \rangle$ is the expectation of momentum, and $\sigma=\sqrt{\left\langle\hat{r}^{2}\right\rangle-\left|\left\langle\hat{r}\right\rangle\right|^{2}}$ is the uncertainty in position with the corresponding conjugate momentum, $p_\sigma$.

Equations-of-motion for the dynamical parameters are easily derived from variation of the time-dependent Schrodinger equation \cite{littlejohn1986,feldmeier2000,grabowskiReviewWP},

\begin{equation}
\mathbf{N} \dot{\mathbf{q}}=\frac{\partial H}{\partial \mathbf{q}}
\end{equation} 
where $H$ is the total energy of the system and $\mathbf{q}$ is the set of all dynamic variables. The norm matrix, $\mathbf{N}$, is defined as follows,

\begin{equation}
N_{a b}=\frac{\partial}{\partial \mathrm{q}_{a}^{*}} \frac{\partial}{\partial \mathrm{q}_{b}} \ln \braket{\Psi^{H}\left(\mathbf{q}^{*}\right)}{\Psi^{H}(\mathbf{q})},
\end{equation}
where $N_{a b}$ is a matrix element of the norm matrix, $\mathrm{q}_{a}$ represents a specific element of the set of time-dependent variational parameters, and $\mathbf{q}_{a}^{*}$ is the complex conjugate of $\mathbf{q}_{a}$. For the single Gaussian basis given in Equation \ref{Eq:GaussianPhys} the equations-of-motion take on a simple Hamilton form \cite{feldmeier2000}.

The total Hamiltonian operator of the  semi-classical many-electron system interacting with classical point-like ions is given by 

\begin{equation}
    \hat{H} = ( \hat{T}_{i} + \hat{V}_{ii} ) + ( \hat{T}_{e} + \hat{V}_{ei} + \hat{V}_{ee} ) + \hat{H}_{Harm} \hspace{3pt},
    \label{Eq:Hamiltonian}
\end{equation} 
where the operators take on their usual classical,

\begin{equation}
    \hat{T}_{i} = \sum_{I} \frac{p_{I}^{2}}{2 M_{I}} \hspace{3pt}, \quad
    \hat{V}_{ii} =  \frac{1}{2} \sum_{I,J} \frac{Z_{I} Z_{J} }{| \vec{R}_{I} - \vec{R}_{J} |} \hspace{3pt}, 
\end{equation} 
and quantum mechanical,

\begin{equation}
    \hat{T}_{e} =  -\frac{1}{2} \sum_{k} \laplacian_{k} \hspace{3pt}, \quad
    \hat{V}_{ei} = \sum_{I}\sum_{k} \frac{-Z_{I}}{\abs{\vec{x}_{k} -\vec{R}_{I}}} \hspace{3pt}, \quad
    \hat{V}_{ee} =  \frac{1}{2}\sum_{k,l}  \frac{1}{\abs{\vec{x}_{k} -\vec{x}_{l}}} \hspace{3pt}, 
\end{equation} 
definitions. Here, $\hat{T}_{e}$ is the electron kinetic energy operator, $\hat{V}_{ei}$ is the electron-ion Coulombic potential energy operator, $\hat{V}_{ee}$ is the electron-electron Coulombic potential energy operator, $\hat{T}_{i}$ is the ion kinetic energy, and $\hat{V}_{ii}$ is the ion-ion Coulombic potential energy. In these expressions, $p_{I}$ represents the classical momentum of the I-th ion, $M_{I}$ is the mass of the I-th ion, $Z_J$ is the number of protons in the J-th ion, and $\vec{R}_{J}$ represents the position of the J-th ion.

The final term in the Hamiltonian defined in Equation (\ref{Eq:Hamiltonian}) is a harmonic energy term ubiquitous throughout WPMD. This term is typically used to constrain the electron's size, which may increase to the point that electron-ion interactions become negligible. For this potential we have used the form suggested by Zwicknagel \emph{et al.} where $H_\mathrm{Harm}=\langle \hat{H}_{Harm} \rangle = \sum_{k=1}^{N_e} \frac{9}{8 \gamma_0^4} \sigma^2$ and $\gamma_0$ is set to to half the simulation box length \cite{Zwicknagel2006,morozov2009}. It should be noted that this term represents a small contribution to the total energy as, in the WDM regime, most wavepackets are constrained enough from the ionic potentials present \cite{davis2020, grabowski2013}. 

\subsection{Extension to Multiple Gaussians}

The primary aim of this work was to investigate the the improvements in describing dense plasmas with WPMD when the basis set is extended to include multiple Guassians. Following the framework of Morozov and Valuev \cite{morozov2012,valuev2015} we extend the basis to include multiple Gaussians as follows,

\begin{equation}
    \psi_{k}(\vec{x}_{k},t)  = n_{k}^{-1/2} \sum_{\alpha=1}^{N_{g}} \varphi_{k \alpha}(\vec{x}_{k},t) \hspace{3pt},
\end{equation} 
where $\psi_{k}$ represents the single particle wave function of the k-th electron, \(N_{g}\) is the total number of Gaussian wave packets per electron, and $n_{k}=\sum_{\alpha, \beta} \int \varphi_{k \alpha}^{*} \varphi_{k \beta} d^{3} x$ is the normalization factor. The term \(\varphi_{k \alpha}\) represents the Gaussian wave packet $\alpha$ in the k-th electron. For multiple Gaussians the simple relationship between the parameters used in Equation (\ref{Eq:GaussianPhys}) and the expectation of the electron physical characteristics is lost. Thus, to simplify the analytic derivation of the energy terms in the Hamiltonian we use the following Gaussian representation, 

\begin{equation}
    \varphi_{k \alpha}(\vec{x}_{k},t) = d_{k \alpha}(t) e^{-a_{k \alpha}(t) (\vec{x}_{k} \cdot \vec{x}_{k}) + \vec{b}_{k \alpha}(t) \cdot \vec{x}_{k} + c_{k \alpha}(t)}.
\end{equation} 
Here, the set of dynamical variables for each Gaussian are \(\mathbf{q}_{k \alpha} = \{ a_{k \alpha}, \vec{b}_{k \alpha}, d_{k \alpha} \}\). These five complex parameters provide a total of ten real dynamic variables for each wave packet. If necessary, these can be mapped directly to the ten physical parameters used in Equation (\ref{Eq:GaussianPhys}). The parameter, \(c_{k \alpha}\), is an not an independent variable and is used to ensure normalization of each Gaussian\cite{valuev2015}. 

Within this framework of multiple Gaussians, the Hartree energy of the system may be expressed as the sum of the ion and electron kinetic energies, along with the electron-ion, electron-electron, and ion-ion potential energies:

\begin{equation}
    H_{H} =  T_{i} + T_{e} +  V_{ii} + V_{ei} +  V_{ee}.
\end{equation} 
In each case, an analytical expression may be derived. For the semi-classical electron terms these are most easily expressed as the product of an overlap integral with a residual. i.e., 

\begin{eqnarray}
   T_{e} &=& \mel**{\Psi^{H}}{\hat{T}_{e}}{\Psi^{H}} = \sum_{k} \sum_{\alpha, \gamma} O_{k \alpha k \gamma} T_{k \alpha k \gamma}^{e}\hspace{3pt}, \\
   V_{ei} &=& \mel**{\Psi^{H}}{ \hat{V}_{ei} }{\Psi^{H}} = \sum_{I} \sum_{k} \sum_{\alpha, \gamma} O_{k \alpha k \gamma} V_{ I k \alpha k \gamma}^{ei} \hspace{3pt}, \\
   V_{ee} &=& \mel**{\Psi^{H}}{ \hat{V}_{ee} }{\Psi^{H}} = \sum_{k,l} \sum_{\alpha,\beta, \gamma, \delta} O_{k \alpha k \gamma} O_{l \beta l \delta} V_{k \alpha l \beta k \gamma l \delta}^{ee} \hspace{3pt},
\end{eqnarray} 
where $O_{k \alpha l \delta}$ represents the overlap between the $\alpha$ Gaussian wave packet of the $k$-th electron with the $\delta$ Gaussian wave packet of the $l$-th electron. Expressions for the three residual terms $T_{k \alpha k \gamma}^{e}$, $V_{ I k \alpha k \gamma}^{ei}$, and $V_{k \alpha l \beta k \gamma l \delta}^{ee}$ and the overlap integral are easily derived in the Gaussian basis and are provided in Ref. \cite{valuev2015}. 

\subsection{Development of a Pairwise Pauli and Correlation Potential}
The Hartree product defined in Equation (\ref{Eq:Hartree}) neglects exchange effects captured by the Slater determinant. Within this approximation, important effects necessary to describe a quantum mechanical system of interacting fermions, such as the Pauli exclusion principle, are neglected. Here we detail the development of a spatially anti-symmetrized pairwise exchange term added between electrons of like spin. This term is equal to the difference between the energy calculated with the Slater determinant and that calculated with a Hartree product. In line with the eFF method, we retain a pairwise Slater determinant to ensure a computational scaling comparable with classical techniques. We construct our pairwise Pauli potential from three terms,

\begin{equation}
    H_{P} =  T_{e}^{P} +   V_{ei}^{P} +  V_{ee}^{P}
\end{equation} 
where,

\begin{eqnarray}
\label{Eq:Pauliterms}
  T_{e}^{P}  &=&  \mel**{\Psi^{S}}{\hat{T}_{e}}{\Psi^{S}} - \mel**{\Psi^{H}}{\hat{T}_{e}}{\Psi^{H}}  \hspace{3pt}, \\
  V_{ei}^{P} &=&  \mel**{\Psi^{S}}{ \hat{V}_{ei} }{\Psi^{S}} - \mel**{\Psi^{H}}{ \hat{V}_{ei} }{\Psi^{H}}  \hspace{3pt}, \\
  V_{ee}^{P} &=&  \mel**{\Psi^{S}}{ \hat{V}_{ee} }{\Psi^{S}} - \mel**{\Psi^{H}}{ \hat{V}_{ee} }{\Psi^{H}} \hspace{3pt},
  \label{Eq:Paulitermsend}
\end{eqnarray} are the Pauli kinetic, Pauli electron-ion potential, and Pauli electron-electron potential energy terms, respectively. Using the usual definition for a two-particle Slater determinant,

\begin{equation}
\label{Eq:SlaterWF}
    \Psi^{S}(\vec{x_{1}}, \vec{x_{2}}) = \frac{1}{\sqrt{2}} [ \psi_{1}(\vec{x}_{1})\psi_{2}(\vec{x}_{2}) - \psi_{1}(\vec{x}_{2})\psi_{2}(\vec{x}_{1}) ] \hspace{3pt},
\end{equation} 
analytic expressions for the two-particle Pauli energy given in Equations (\ref{Eq:Pauliterms}-\ref{Eq:Paulitermsend}) were derived. 

\begin{figure}[t]
    \includegraphics[width=1.0\linewidth]{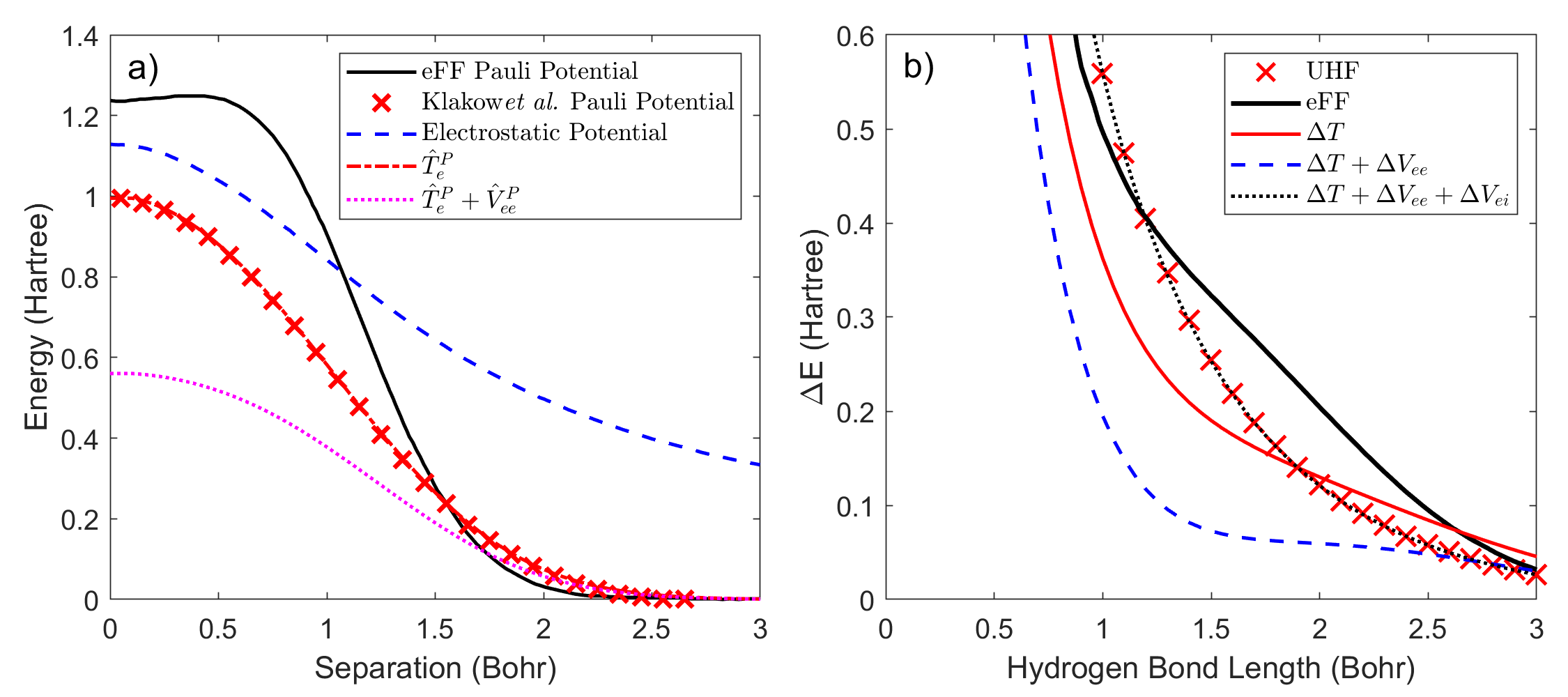}
    \caption{\textbf{Validation of the three-term Pauli potential.} (a) The Pauli potential consisting of all three energy terms is compared to the potential used in eFF and that developed by Klakow \emph{et al} which uses only the kinetic energy term \cite{su2009}. Results are given for the exchange energy between two electrons of fixed-width $\sigma=\sqrt{3/4}$. Unlike the electrostatic energy, shown by the dashed line, the Pauli potentials do not exhibit long-ranged behaviour. (b) Verification of the Pauli potential following the method suggested by Xiao \emph{et al.} \cite{xiaothesis2015}. The different energy contributions upon anti-symmetrization for the triplet $\text{H}_{2}$ system. In this system, the unrestricted Hartree-Fock (UHF) result is exact. Each electron is represented by a single Gaussian of radius 1.7~Bohr centered on the atom. Note, the eFF potential does not match the exact UHF result.}
    \label{fig:figure2_Ex}
\end{figure}

The majority of WPMD techniques that do not implement full exchange have opted to use a Pauli potential based solely on the kinetic energy component of the Pauli exchange \cite{boal1988,klakow1994,knaup2001,su2009}. In addition, while some authors retain the dependence on electron momentum \cite{klakow1994}, many models, including eFF, simplify the terms further by neglecting this dependence \cite{boal1988, su2007}. Figure \ref{fig:figure2_Ex}a compares our pairwise Pauli potential with other published results. The results by Klakow \emph{et al.} contain only the kinetic energy contribution to exchange and, for this system, agree with our $T_{e}^{P}$ term. The eFF model also makes use of the kinetic energy Pauli potential but differs from Klakow \emph{et al.} due to the incorporation of empirical scaling parameters \cite{su2007}. Due to the lack of ions in the system, our total Pauli term contains just one additional term, $V_{ee}^{(P)}$. This term acts to lowers the exchange energy when compared to both the Klakow potentials. The eFF potential, which lies above the Klakow model, performs quite poorly for this system. This is unsurprising as, with a simple form, it is attempting to correct for several approximations in the model.

In Figure \ref{fig:figure2_Ex}b, we compare the energy of the anti-bonding of the $\text{H}_{2}$ molecule with that predicted by unrestricted Hartree Fock (UHF). For the anti-bonding case, there is no correlation, which makes it the most straightforward system to examine exchange calculations. It is clear that the kinetic energy term is dominant and primarily responsible for the Pauli exclusion principle; however, the other terms are not negligible and play an important role in stabilizing molecules and preventing Gaussian coalescence \cite{xiaothesis2015}. When all three Pauli energy terms are added together, the result matches the exact UHF result and establishes our Pauli potential's validity. Note that the eFF potential, denoted by the solid black line, does not match the exact UHF result. 

One of the reasons for the success of the eFF technique over previous models is the inclusion of additional scaling parameters in the Pauli potential. These parameters, matched to a set of molecular test structures, account for the lack of full exchange, the limited basis set, and the neglect of correlation, where we define correlation as the difference between the exact energy and the Hartree-Fock (HF) energy. Despite our expanded basis, two approximations persist in our model; these are pairwise exchange and lack of correlation. The inclusion of correlation in such models is difficult and decades old problem in quantum mechanical many-body systems \cite{lowdin1955,lavrinenko2019,chachiyo2020,lavrinenko2021}. Motivated by eFF, we used a valence bond (VB) wave function to develop a simple pairwise correction term to account for these deficiencies. The approximation takes on a similar form as the Pauli potential,

\begin{equation}
    H_{C} =  T_{e}^{C} +   V_{ei}^{C} +  V_{ee}^{C} \hspace{3pt},
\end{equation} 
where

\begin{eqnarray}
\label{Eq:Corterms}
  T_{e}^{C}  &=&  \mel**{\Psi^{VB}}{\hat{T}_{e}}{\Psi^{VB}} - \mel**{\Psi^{H}}{\hat{T}_{e}}{\Psi^{H}}  \hspace{3pt}, \\
  V_{ei}^{C} &=&  \mel**{\Psi^{VB}}{ \hat{V}_{ei} }{\Psi^{VB}} - \mel**{\Psi^{H}}{ \hat{V}_{ei} }{\Psi^{H}}  \hspace{3pt}, \\
  V_{ee}^{C} &=&  \mel**{\Psi^{VB}}{ \hat{V}_{ee} }{\Psi^{VB}} - \mel**{\Psi^{H}}{ \hat{V}_{ee} }{\Psi^{H}}. 
  \label{Eq:Cortermsend}
\end{eqnarray} 
As before, utilizing the usual definition of the two-particle VB wave function, 

\begin{equation}
    \Psi^{VB}(\vec{x_{1}}, \vec{x_{2}}) = \frac{1}{\sqrt{2}} [ \psi_{1}(\vec{x}_{1})\psi_{2}(\vec{x}_{2}) + \psi_{1}(\vec{x}_{2})\psi_{2}(\vec{x}_{1}) ] \hspace{3pt},
\end{equation} 
we derived analytic expressions for the pairwise VB correction energy given in Equations (\ref{Eq:Corterms}-\ref{Eq:Cortermsend}). The VB correction was implemented between pairs of electrons with the same spin. Opposite spin electrons were assumed to be well separated spatially, due to the Pauli potential, and to not contribute significantly to this term \cite{suthesis2007}.

The total energy of our system, $\text{H} = \text{H}_{H} + \text{H}_{P} + \text{H}_{C} $ may be written as follows,

\begin{equation}
    H = (T_{i} + V_{ii}) + ( T_{e} + V_{ei} + V_{ee} ) + ( T_{e}^{P} + V_{ei}^{P} + V_{ee}^{P} )\delta_{\uparrow \uparrow} + \rho ( T_{e}^{C} + V_{ei}^{C} + V_{ee}^{C} )\delta_{\uparrow \downarrow} \hspace{3pt},
\end{equation} 
where $\delta_{\uparrow \uparrow}$ equaled unity when the spins of the electrons were parallel and zero otherwise, while $\delta_{\uparrow \downarrow}$ was set to zero when the spins of the electrons are parallel and unity when anti-parallel. The parameter, $\rho$, must be chosen \emph{a-priori} for the system as interest. For a hydrogen molecule, a value of $\rho=1$ approaches the exact potential energy surface. For calculating the pressure of dense hydrogen, we scaled this value to match experimental results. We note that the size of $\rho$ as $N_g\rightarrow\infty$ gives an indication of the remaining approximations in the model.

\subsection{Implementation of Periodic Boundary Conditions}

The simulation of dense plasmas necessitates the use of a finite size simulation box utilizing periodic boundary conditions; this brings with it additional challenges. The first is the evaluation of long-range forces associated with electrostatic interactions. We utilize the same scheme as eFF where the electrostatic energies are multiplied by a seventh order spline that goes from 1 to 0 over a radial distance $r_{cut}$, defined so that the first, second, and third derivatives at the endpoints are zero \cite{izvekov2008,su2009}. i.e.,

\begin{equation}
f_{\text {cutoff }}=20 x^{7}-70 x^{6}+84 x^{5}-35 x^{4}+1 \hspace{3pt},
\end{equation} 

where $x=x/r_\mathrm{cut}$. The cutoff range, $r_\mathrm{cut}$, is chosen to be equal to half the simulation box length. This technique enables the use of the minimum-image convention, retaining one of the most desirable properties of WPMD, which is the ability to simulate large systems of particles. Finally, as demonstrated in Figure \ref{fig:figure2_Ex}a, the newly defined exchange terms do not exhibit the same long-range characteristics as the electrostatic terms and thus are not multiplied by the spline.

When applying the minimum image convention to the electrons, the Gaussians that comprise that electron must be treated as a single particle; that is to say, they must be shifted together. Thus, to apply the minimum image convention, we use the expectation of each electron's position. Suppose care is not taken, and the Gaussians are individually shifted. In that case, the unphysical situation where a particle interacts with the same electron on both the left and the right can arise.  

Finally, the Pauli and VB correction electron-ion energy terms are not genuine pairwise terms. Each term in the summation involves two electrons plus an ion, essentially making it a three-body potential. To ensure consistent calculation of this term, we periodically shift the ion location to position it within half of the box length of the mid-point between the two electrons. We note that this differs from consistently shifting the ion towards the electron only when the two electrons are spatially separated; in such situations, the Pauli and VB correction energy terms are negligible. 

%%%%%%%%%%%%%%%%%%%%%%%%%%%%%%%%%%%%%%%%%%%%%%%%%%%%%%%%%%%%%%%%%%%%%%%%%%%%%%%%%%%%%%%%%%%%%%%%%%%%%%%%%%%%%%%%%%%%%%%%%%%%%%%%%%%%%%%%%%%%%%%%%%%%%%%%%%%%%
%%%%%%%%%%%%%%%%%%%%%%%%%%%%%%%%%%%%%%%%%%%%%%%%%%%%%%%%%%%%%%%%%%%%%%%%%%%%%%%%%%%%%%%%%%%%%%%%%%%%%%%%%%%%%%%%%%%%%%%%%%%%%%%%%%%%%%%%%%%%%%%%%%%%%%%%%%%%%

\section{Results}

Hydrogen plasmas are one of the simplest physical systems and, for this reason, have become the prototypical test-bed for atomistic models of dense plasmas. For example, the hydrogen atom has a well-defined analytical solution. Figure \ref{fig:Hatom} demonstrates that we approach this analytic solution with an increasing Gaussian basis. The ground state energy can be obtained to within a percent, utilizing a four Gaussian basis. Beyond this number, we observe only minor changes in the ground state energy and the electron wavefunction shape. It should also be noted that the most significant improvement to both occur when increasing the basis from one to two Gaussians. 

\begin{figure}[H]
    \includegraphics[width=0.95\linewidth]{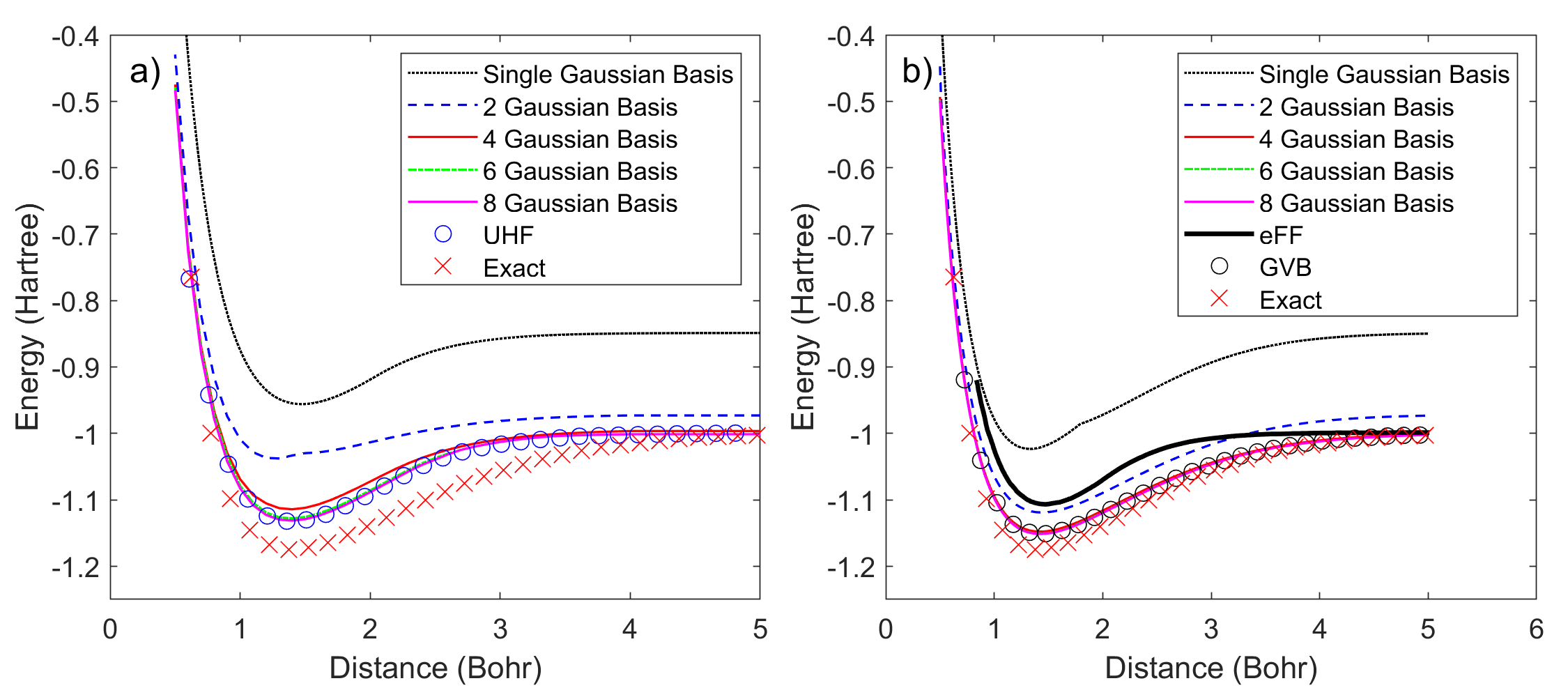}
    \caption{\textbf{Binding energy of the $\mathbf{H_{2}}$ molecule.} (a) VB correction is excluded. Circles represent the unrestricted Hartree-Fock (UHF) solution and crosses represent the exact solution. (b) VB correction ($\rho=1$) is included. Circles represent the generalized valence bond (GVB) solution and crosses denote the exact solution. The thick solid line represents the solution obtained using the eFF method.}
    \label{fig:figure3_H2}
\end{figure}

We now turn our attention to modeling the energy of the hydrogen molecule. In Figure \ref{fig:figure3_H2} we plot the potential energy surface of $\text{H}_{2}$ and show improvement in representation as the Gaussian basis is increased. Figure \ref{fig:figure3_H2}a displays the potential energy curve calculated without including the VB correction; in this case, the binding energy approaches the unrestricted Hartree-Fock (UHF) result. In this case, negligible improvement is observed beyond six Gaussians, slightly more than needed to describe the hydrogen atom accurately. Figure \ref{fig:figure3_H2}b displays the potential energy curve calculated with an increasing number of Gaussians per electron, this time with the VB correction ($\rho=1$). In this case, as the number of Gaussians is increased, the potential energy curve tends toward the generalized valence bond (GVB) result \cite{suthesis2007}, validating the implementation of the multiple Gaussian basis. The eFF results, also shown in Figure \ref{fig:figure3_H2}b, lie between the one and two Gaussian lines, clearly indicating that the scaling factors in eFF somehow address the impediment from the limited basis and lack of correlation. 

The potential energy curves for the two Gaussian case in Figure \ref{fig:figure3_H2}a, and one Gaussian case in Figure \ref{fig:figure3_H2}b, both exhibit a kink around 1.6~Bohr. We attribute this to the restricted basis attempting to represent two phenomenologically different configurations to the left and right of the kink; that is, the expectation of the two electrons' position being located between the two ions when close together and centered near the ions when separated. This effect is less apparent when a larger basis is used. 

\begin{figure}[H]
    \centering
    \includegraphics[width=0.6\linewidth]{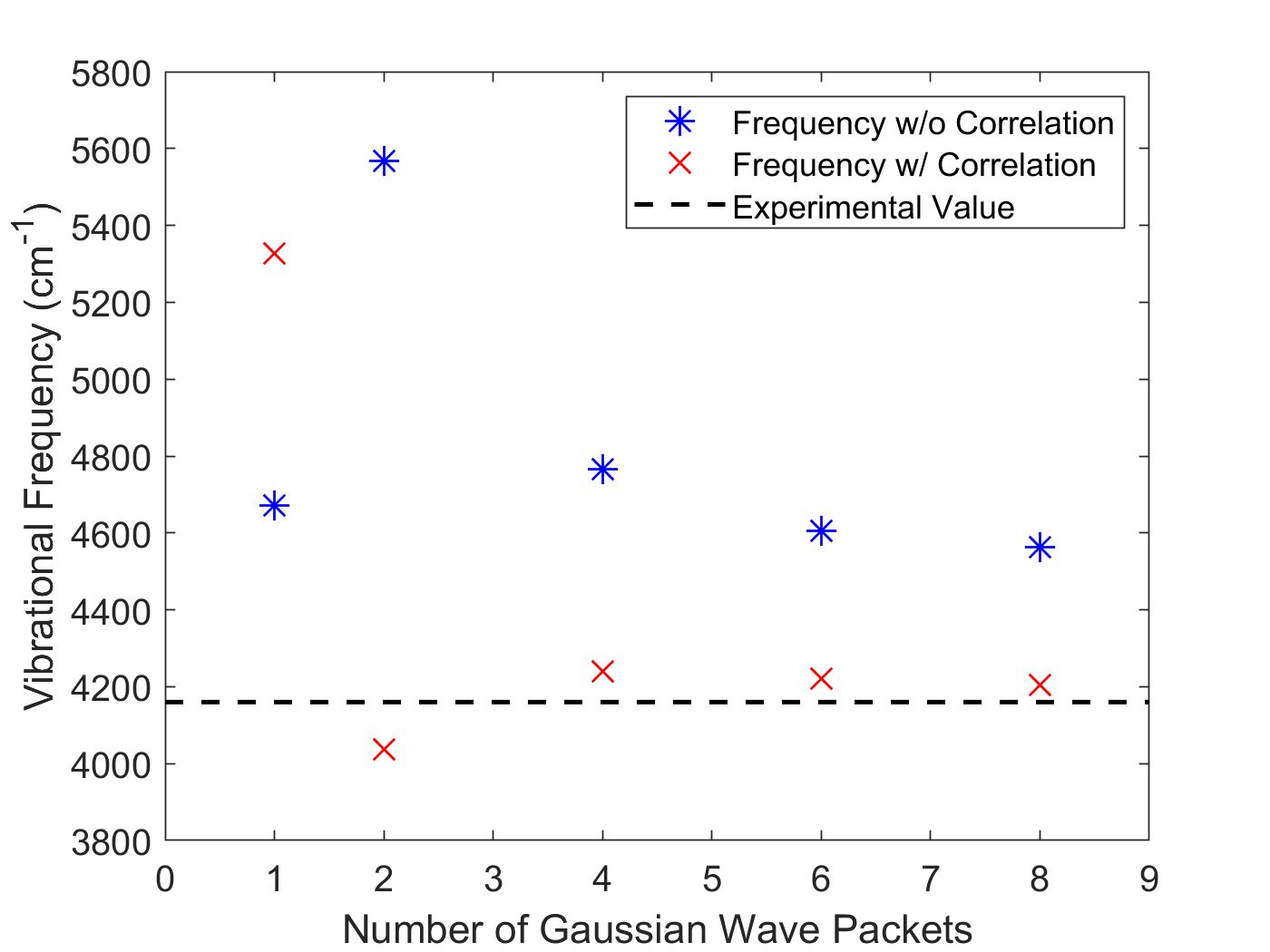}
    \caption{\textbf{Vibrational frequency of the $\mathbf{H_{2}}$ molecule with an increasing Gaussian basis set.} Blue stars are the frequency with no VB correction. Red crosses are the freqeucnies calculated with the VB correction ($\rho=1$). The black dotted line represents the experimental value of the fundamental frequency of $\text{H}_{2}$. In each case the fundamental frequency was calculated by displacing two H ions by 0.01~Bohr and observing the resulting oscillations.}
    \label{fig:figure4}
\end{figure}

Figure \ref{fig:figure3_H2} represents the static use of the code. To validate the dynamic component, We calculated the fundamental vibrational frequency of $\text{H}_{2}$ through the dynamical equations-of-motion. In the simulations, the two hydrogen nuclei we separated by a small distance of 0.01~Bohr and released, the time step was 0.603~as, and the total run time was 73~fs. The energy was conserved to within \(10^{-12}\)~Hartree. The $V_{ii}$ energy was plotted versus time, creating a sinusoidal graph where the peak to peak time was averaged over ten oscillations to obtain the fundamental vibrational frequency. As we increased the number of Gaussians used, the results tended towards a consistent result. Without the VB correction, this was found to be 4563~cm$^{-1}$, almost 10~\% higher than the experimental value of 4161~cm$^{-1}$ \cite{dickenson2013}. In the simulation in which correlation was accounted for, the frequency was found to be 4204~cm$^{-1}$, just 1\% higher than the experimental value. As before, a significant improvement was found when the number of Gaussians was increased from one to two. 

\begin{figure}[H]
    \centering
    \includegraphics[width=1.0\linewidth]{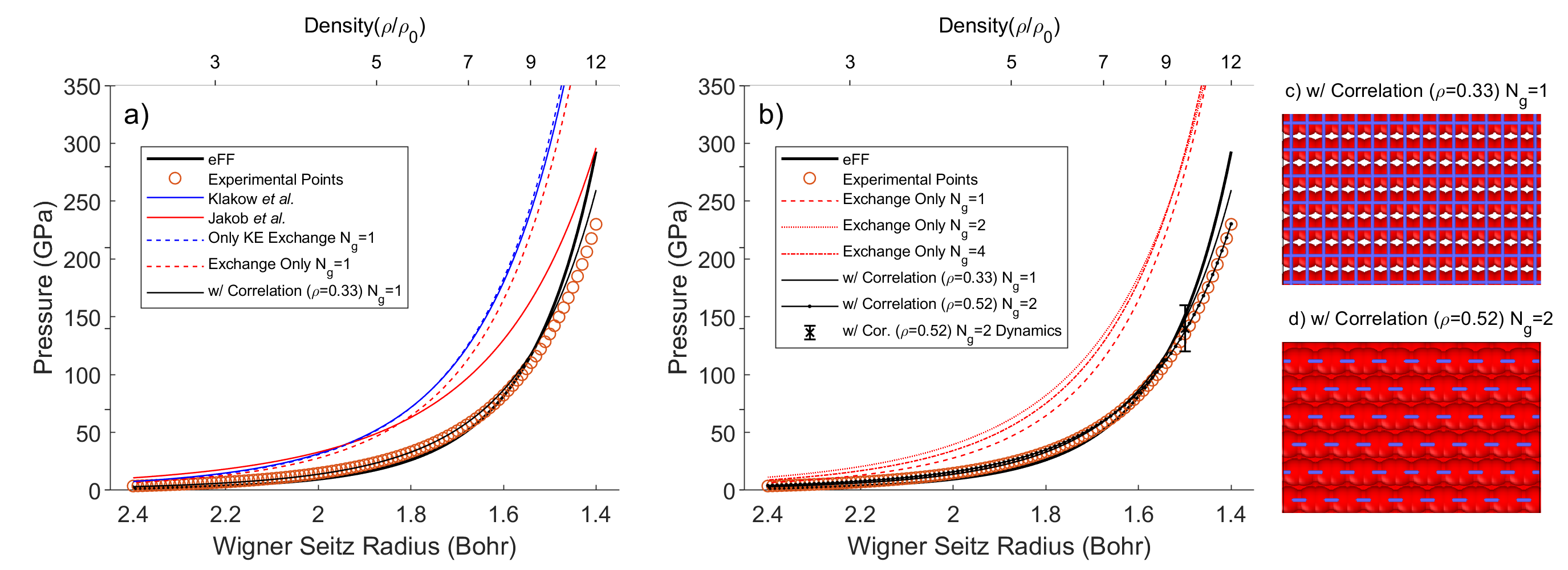}
    \caption{\textbf{The Pressure of Dense hydrogen.} (a) A comparison of our results to results to those in the literature for a single Gaussian basis. Our results were calculated at 0~K while the published results, both experimental and theoretical, were calculated at 300~K. The small offset between the Klakow {et al.} results and our results using purely the kinetic energy component of exchange demonstrate the applicability of this approximation. (b) Extension to multiple Gaussians. The red lines, calculated with pairwise exchange only, show that increasing the number of Gaussians has only a small effect on the pressure curve. The VB correction is needed to best match experimental results. (c) Orthorhombic crystal structure of hydrogen at Wigner-Seitz radius of 1.6~Bohr for a single-Gaussian basis and scaled VB correction ($\rho=0.33$). (d) Hexagonal-close-packed crystal structure of hydrogen at a Wigner-Seitz radius of 1.6 with a two Gaussian basis and scaled VB correction ($\rho=0.52$). }
    \label{fig:figure5}
\end{figure}

We have demonstrated the applicability of the method to the hydrogen trial systems presented thus far. We now turn our attention to modeling a periodic hydrogen plasma system. For this work, we utilized systems consisting of 1024 ions and an equal number of electrons. At this time, the code is implemented serially in MATLAB and is unable to be run for long times across all permutations of approximations and basis sets. Thus, for the majority of results presented in Figure \ref{fig:figure5}, we compare the pressure of an energy-minimized system at 0~K, calculated according to the virial theorem \cite{landau}, to experimental results and other models calculated at 300~K. However, at these densities, the difference in pressure between 300~K and 0~K is negligible. This was confirmed by comparing, in Figure \ref{fig:figure5}a, the data from Klakow \emph{et al.}  (solid blue line) and our kinetic energy exchange (dashed blue line), both calculated with a single Gaussian basis; they are almost identical over the range of densities considered. 

In Figure \ref{fig:figure5}a, which shows only results calculated with a single-Gaussian basis, we find that results calculated with our pairwise exchange term diverge from those of Jakob \emph{et al.} which were calculated using exact exchange. However, with the addition of the VB correction, in this case with the scaling parameter $\rho=0.33$, we are able to better match experimental results obtained on diamond anvil cells \cite{Loubeyre1996}. In this case, the VB correction is accounting for the pairwise exchange, the limited basis, and lack of correlation. We note that our result follows more closely the experimental data than eFF. 

In Figure \ref{fig:figure5}b the effect of an increased basis is shown for the same dense hydrogen system, both with and without the VB correction. Interestingly, simply with the pairwise exchange energy, an increased Gaussian basis does not significantly improve the results. This suggests the system's limitation can be traced to either the lack of correlation or the pairwise exchange. However, with the VB correction, we can scale $\rho$ to lower the pressure. With a two Gaussian basis and a scaled VB correction ($\rho=0.52$), we can exactly match the experimental results across the density regime tested. These results suggest that improved correlation and full exchange play a larger role than the expanded basis. With that said, if we investigate the crystal structure of the two minimized systems with the correction, we find the single-Gaussian basis has an orthorhombic structure (c.f. Figure \ref{fig:figure5}c). In contrast, the two-Gaussian system has the correct hexagonal-close-packed structure (c.f. Figure \ref{fig:figure5}d). 

For our best model, the two-Gaussian basis with $\rho=0.52$, we were able to run a periodic dynamical simulation using a 0.24~as timestep for 130~as. During this time, we scaled the ion velocities to have a kinetic energy consistent with 300~K and let the electrons equilibrate with the ions. During this simulation, the pressure was observed to oscillate between 125~GPa and 160~GPa, denoted by the cross and error bars in Figure \ref{fig:figure5}b. This simulation demonstrates the model's applicability but highlights the importance of parallelizing the code for modeling larger systems.

%%%%%%%%%%%%%%%%%%%%%%%%%%%%%%%%%%%%%%%%%%%%%%%%%%%%%%%%%%%%%%%%%%%%%%%%%%%%%%%%%%%%%%%%%%%%%%%%%%%%%%%%%%%%%%%%%%%%%%%%%%%%%%%%%%%%%%%%%%%%%%%%%%%%%%%%%%%%%
%%%%%%%%%%%%%%%%%%%%%%%%%%%%%%%%%%%%%%%%%%%%%%%%%%%%%%%%%%%%%%%%%%%%%%%%%%%%%%%%%%%%%%%%%%%%%%%%%%%%%%%%%%%%%%%%%%%%%%%%%%%%%%%%%%%%%%%%%%%%%%%%%%%%%%%%%%%%%

\section{Discussion}

Atomistic simulations such as WPMD that go beyond the BO approximation may be necessary to describe the dynamics of dense plasmas systems and, in particular, non-equilibrium matter. However, in many implementations of WPMD, multiple approximations are used to achieve computational efficiency. This efficiency has led to the widespread use of the eFF flavor of WPMD, which, despite the restrictive single-Gaussian basis, has achieved remarkable success. This can, in part, be attributed to the semi-empirical scaling parameters that simultaneously attempt to correct for deficiencies in the basis, approximations to exchange, and lack of correlation. However, the use of scaling parameters makes the method's applicability in untested regimes, such as for higher-Z or extremely dense plasmas, questionable. 

To reduce the number of scaling parameters, we have implemented a version of WPMD with an extended multiple Gaussian basis to describe the behavior of periodic systems of dense hydrogen. Furthermore, we use the improved Pauli potential suggested by H. Xiao, which includes the electron-electron and electron-ion components of exchange. However, in order to achieve computational scaling comparable to classical methods, one of the distinct advantages of the eFF method, we retain the use of a pairwise Pauli potential. We find improvements in the description of hydrogen atoms, molecules, and plasma systems as the number of Gaussians are increased from one to four, in agreement with previous work on non-periodic systems of a few electrons \cite{grabowski2013, morozov2012,valuev2015}. Notably, increasing the basis from one to two Gaussians provides the greatest improvement.

With the improved basis and exchange, the most significant remaining error is attributed to the model's lack of correlation. We implemented a simple VB correction based on the VB wave function, which we demonstrate in the hydrogen molecule and plasma. However, to match experimental results, we must utilize a single-scaling parameter on this correction term. Future work will focus on improving the correlation part of this term, which could be modified to take into account local order \cite{xiaothesis2015}, or exploiting the robust correlation functionals within DFT \cite{lavrinenko2019}. In addition, we are working on parallelization of the code, which will allow for the description of larger plasma systems for greater times.

%%%%%%%%%%%%%%%%%%%%%%%%%%%%%%%%%%%%%%%%%%

\authorcontributions{
W.A.A. and T.G.W. contributed equally to this manuscript.
}

%%%%%%%%%%%%%%%%%%%%%%%%%%%%%%%%%%%%%%%%%%

\funding{This material is partially based upon work supported by the U.S. Department of Energy, Office of Science, Office of Fusion Energy Sciences under Grant No. DE-SC0019268.}

%%%%%%%%%%%%%%%%%%%%%%%%%%%%%%%%%%%%%%%%%%
\acknowledgments{Special thanks to Nuvraj Bilkhu, Jacob Molina, and Cameron Allen}

%%%%%%%%%%%%%%%%%%%%%%%%%%%%%%%%%%%%%%%%%%
\conflictsofinterest{The authors declare no conflict of interest. The funders had no role in the design of the study; in the collection, analyses, or interpretation of data; in the writing of the manuscript, or in the decision to publish the results. }

%%%%%%%%%%%%%%%%%%%%%%%%%%%%%%%%%%%%%%%%%%
%=====================================
% References
%=====================================
\reftitle{References}

\externalbibliography{yes}
\bibliography{References}

\end{document}